\documentclass[10pt]{article}
\usepackage{latexsym}
\usepackage[dvips]{graphicx}
\usepackage{overcite}
\usepackage{amssymb}
\usepackage{amsbsy}
\usepackage{amsmath}
\usepackage{mathrsfs}
\usepackage{xr}
\usepackage{booktabs} 
\usepackage{multirow} 
\usepackage{xcolor}	  
\usepackage{mathtools}
\usepackage{graphicx}
\usepackage{tabularx}
\usepackage[labelformat=simple, font=small,labelfont=bf]{caption}
\usepackage{placeins}
\usepackage{authblk}
\graphicspath{ {./plot/} }

\DeclareRobustCommand*{\Figure}[3]{
   \begin{figure}[!htb]
   \begin{center}
   \noindent
   \includegraphics[width=#2]{#1}  
   \end{center}
   \caption{#3}
   \addtocontents{lof}{\vspace{\baselineskip}}
   \label{fig:#1}
   \end{figure}
}

\newcommand{\be}{\begin{equation}}
\newcommand{\ee}{\end{equation}}

\begin{document}

\title{Micellization of Diblock Copolymer Modified by Cononsolvency Effect}

\author[1]{Xiangyu Zhang\footnote{E-mail: xzhan357@jh.edu}}
\author[2]{Jing Zong}
\author[3]{Dong Meng}
\affil[1]{Department of Chemical and Biomolecular Engineering, John Hopkins University, Baltimore, MD 21218, United States}
\affil[2] {Dave C. Swalm School of Chemical Engineering, Mississippi State University, MS 39762, United States}
\affil[3]{Biomaterials Division, Department of Molecular Pathobiology, New York University, New York, NY 10010, United States}

\maketitle

\begin{abstract}
Spherical micelle modified by cononsolvency effect is investigated by using self-consistent field theory (SCFT) and random phase approximation, which is formed by diblock copolymers consisted of one permanent hydrophobic block and one hydrophilic block. The addition of the cosolvent will bring about the cononsolvency effect on the hydrophilic block. The result shows that cononsolvency effect will expand the micelle size and changes the critical micelle concentration. By analyzing density profiles predicted by SCFT calculation, the micelle shell exhibits extension-collapse-extension transition with the addition of the cosolvent, which is responsible for the micelle size change. The driving force of the micellization is analyzed in the framework of SCFT calculation. The conventional micelle, which is formed by amphiphilic diblock copolymer in pure solvent, is compared with the micelle modified by cononsolvency effect. The reduction of the core-block - solvent contact area drives the formation of the conventional micelle. But in cononsolvency modified micelle, it is shown that the shell-block - cosolvent favorable interaction also plays the role to minimize the total free energy, whose mechanism is significantly different from that of the conventional micelle.

\end{abstract}

\section{Introduction}
It is well-known that abundant structures can be formed by the self-assembly of the amphiphilic diblock copolymer in the solvent, which is consisted of one permanent hydrophobic block and one hydrophilic block.\cite{alexandridis2000amphiphilic,mai2012self} Among all structures formed by the amphiphilic diblock copolymer, the most extensively studied one is the spherical micelle, and it has a range of applications, such as, catalysis, drug delivery, biological imaging, \textit{etc}.\cite{ge2007stimuli,riess2003micellization,kabanov2002pluronic,maysinger2007fate} The micelle morphology of the amphiphilic diblock copolymer is majorly influenced by the following factors, solvent compositions, copolymer composition and concentration, presence of additives, if we do not consider external stimulation.\cite{mai2012self,zhang1996multiple,zhang1998formation} However, the use of the solvent composition to stimulate micro-structure change has been a tricky problem as some "counter-intuitive" or intriguing behaviors related to cosolvency or cononsolvency effect may arise when multi-solvents are involved, for which our understanding is still lack. \cite{rao2007cononsolvency,wang2008investigation,michailova2010nanoparticles,kyriakos2016quantifying,geiger2021pmma,ko2021co} 
.


\par
Cosolvency means that the mixture of two poor solvents can enhance the polymer solubility. \cite{dudowicz2015communication} The common example system is poly(methyl methacrylate)(PMMA) immersed in water+alcohol mixture, which is a UCST type polymer. \cite{cowie1987alcohol} Water and alcohol are both non-solvents for PMMA. But the mixture of them can enhance the solubility at intermediate composition range, corresponding to the decrease of the critical temperature.\cite{cowie1987alcohol} Cononsolvency means that the mixture of two good solvents can create a bad solution condition for the polymer, decreasing the solubility. \cite{dudowicz2015communication} The common example system is PNIPAm immersed in water+alcohol mixture, which is a LCST type polymer. The significant collapse of the PNIPAm chain can be observed at intermediate solvent composition range. \cite{wang2012conformational, zhang2001reentrant} Obviously, the occurrence of cosolvency effect will destroy the micro-structure formed by diblock copolymer, as the increase of the solubility will just dissolve the polymer.\cite{ko2021co} Accordingly, we should focus on the various behaviors of micelle brought by cononsolvency effect.
\par
In general, the cononsolvency effect on the micelle formed by diblock copolymers can be categorized into two types. One is micelle formation driven by cononsolvency, in whose system micelle is composed of double hydrophilic block polymers.\cite{rao2007cononsolvency,ge2007stimuli,michailova2010nanoparticles,michailova2018self,wu2022synthesis} For example, poly(N-isopropylacrylamide)-b-poly(oligo(ethylene glycol) methyl ether methacrylate)(PNIPAM-b-POEGMA) diblock copolymers are unimers in pure water or pure methanol below LCST. But PNIPAm-core micelle can be observed in methanol/water mixture.\cite{rao2007cononsolvency} The other type is micelle morphology modified by cononsolvency, in which micelle is composed of one permanent hydrophobic block and one hydrophilic block, and the hydrophobic block constitutes as the micelle core. \cite{kyriakos2014cononsolvency,wang2008investigation,kyriakos2016quantifying,zhou2016pfs,ko2021co} For example, the shell structure of the spherical micelle formed by poly(methyl methacrylate)‑b‑poly(N‑isopropylacrylamide)(PMMA-b-PNIPAM) will become shrank with the addition of methanol in water-rich region (not in PMMA cosolvency region).\cite{ko2021co} In this study, we focus on the second case. 
\par
Thereby, the mechanism of cononsolvency should first be discussed, though it is still under debate. Bharadwaj \textit{et al.} categorized current proposed driving force for cononsolvency into four aspects, (a) cosolvent-solvent attraction, (b) enthalpic bridging, (c) geometric frustration, (d) cosolvent surfactant mechanism.\cite{bharadwaj2022cononsolvency} Full details can be found in that paper, which will not be elaborated here. The important point is that (b)(c)(d) actually can be generalized as one effect, which is the strength of polymer-cosolvent affinity force, no matter whether it is driven by entropy or enthalpy.\cite{zhang2020unified, zhang2024general} So, the mechanism of cononsolvency becomes the question whether cosolvent prefer more contacting with polymer or more with solvents, corresponding to P-C driven or S-C driven mechanisms, respectively. \cite{zhang2020unified, zhang2024general} For cononsolvency modified or driven micelle, both blocks can be the core if cononsolvency effect was driven by S-C attraction, but that is not the case in experiments. Accordingly, the model to study cononsolvency effect on micelle behaviors should be built up based on P-C attractive interaction. 
\par
The manuscript is organized as following. In the first part, Random phase approximation is used to predict phase instability boundary. In the second part, SCFT calculation provides information about morphology change upon the addition of cosolvents with different quality. In the last part, thermodynamics information about cononsolvency modified micelle is presented, and it is compared with conventional micelle in single solvent.
\par
\section{Model and Method}
We consider a system containing solvents(S), cosolvents(C) and diblock copolymer chains(A-b-B) with length of each block being $N_{A}=32$ and $N_{B}=16$ at temperature $T$ in volume $V$. The chain length of the polymer is $N_{P}=N_{A}+N_{B}$. Non-bonded potential is described by Flory-Huggins $\chi$, and bonded potential is given by discrete gaussian bond. B-blocks have the strong repulsion to solvents, becoming the micelle core, and A-block exhibits cononsolvency effect by tuning A-block - cosolvents attractions. The following subsection gives the detailed description about random phase approximation and self-consistent field method derivation.  
\subsection{A-B/S/C System Random Phase Approximation} \label{method:RPA}
Random Phase Approximation developed by Leibler\cite{leibler1980theory} is applied. The average density over the whole system of $i$ component can be defined by, $\langle\rho_{i}({\bf r})\rangle=f_{i}$. The order parameter describing the fluctuation of position ${\bf r}$ is defined as, $\Psi_{i}({\bf r})=\langle\rho_{i}({\bf r})-f_{i}\rangle$. Following the standard procedure, the order parameter can be transformed to wave vector space and be expressed as a function of external potential field,
\begin{equation}
\Psi_{i}({\bf q})=-\beta\sum_{j}^{A,B,S,C}\tilde{S}_{ij}({\bf q})U_{j}({\bf q})=-\beta\sum_{j}^{A,B,S,C}S_{ij}({\bf q})U_{j}^{eff}({\bf q})
\end{equation}
where $\beta=1/(k_{B}T)$, $k_{B}$ is the Boltzmann constant and $T$ is the temperature, ${\bf q}$ is the wave vector, $U_{i}^{eff}({\bf q})=U_{i}({\bf q})+\sum_{j}^{j\neq i}V_{ij}\Psi_{j}({\bf q})+V$, $V_{ij}=k_{B}T\chi_{ij}$, $U_{i}$ is the $i$ component potential field, $V$ is the excluded volume effect, $\chi_{ij}$ is the interaction strength between $i$ and $j$ component, and $S_{ij}$ is the ideal state structure factor, $\tilde{S}_{ij}$ is the structure factor under the external potential field. Ideal state structure factor is already known and can be expressed as following, 
\begin{equation}
\begin{gathered}
S_{AS}=S_{BS}=S_{AC}=S_{BC}=0  \qquad \text{(no connectivity)}\\
S_{SS}=S_{CC}=1  \qquad \text{(unit length)}\\
S_{AA}=N_{P}\phi_{P}(\frac{2}{x^{2}}(fx+\exp(-fx)-1))\\
S_{BB}=N_{P}\phi_{P}(\frac{2}{x^{2}}((1-f)x+\exp(-(1-f)x)-1))\\
S_{AB}=S_{BA}=\frac{1}{2}N_{P}\phi_{P}(\frac{2}{x^{2}}(x+\exp(-x)-1))-\frac{1}{2}S_{AA}-\frac{1}{2}S_{BB}\\
\end{gathered}
\end{equation}
where $N_{P}$ is the A-B diblock copolymer chain length, $f$ is the fraction of the A-block, $\phi_{P}$ is the polymer volume fraction, and $x\equiv q^{2}R_{g}^{2}$, $R_{g}^{2}$ is the radius of gyration. \\
Four $\Psi_{i}$ expressions as a function of $S_{ij}$ combined with the incompressibility condition $\Psi_{A}+\Psi_{B}+\Psi_{S}+\Psi_{C}=0$ can be solved with five unknown variables, which are $\Psi_{A}$, $\Psi_{B}$, $\Psi_{S}$, $\Psi_{C}$ and $V$. By using the right-side equality in Eq. (1), $\tilde{S}_{ij}$ can be obtained. The diverging behavior of $\tilde{S}_{AB}$ (denominator touching 0) indicates the phase separation. And the reciprocal of corresponding wave vector length $x$ where diverging occurs can tell the phase separation length scale. So, the solution of two equations, the denominator equal to 0 and the first order derivative of denominator equal to 0, can locate the critical polymer concentration ($\phi_{P}^{*}$) and the corresponding phase transition length scale($x^{*}$).
\begin{equation}
\begin{gathered}
\mathrm{denominator}[\tilde{S}_{AB}(\phi_{P},x)]=0 \\
\frac{\partial \mathrm{denominator}[\tilde{S}_{AB} (\phi_{P},x)]}{\partial x}=0\\
\end{gathered}
\end{equation}

\subsection{Self-Consistent Field Theory}
The partition function ($\mathcal{Z}$) in grand canonical ensemble of A-B/S/C system can be written in the form,
\begin{equation}
\begin{split}
{\Xi}(\mu_{P},\mu_{S},\mu_{C},V,T) = & \sum_{n_{P}=0}^{\infty} \sum_{n_{S}=0}^{\infty} \sum_{n_{C}=0}^{\infty} \lambda_{T}^{-3n_{P}N-3n_{S}-3n_{C}} e^{\mu_{P}n_{P}+\mu_{S}n_{S}+\mu_{C}n_{C}} \frac{1}{(n_{\rm P})! n_{\rm S}! n_{\rm C}!} \\
& \prod_{j=1}^{n_{\rm S}}\int {\rm{d}} {{\bf r}_{{\rm S},j}} \prod_{j'=1}^{n_{\rm C}}  \int {\rm{d}} {{\bf r}_{{\rm C},j'}}\prod_{k=1}^{n_{\rm P}}  \prod_{s=1}^{N_P}  \int {\rm{d}} {{\bf R}_{k,s}}  \exp \left( - \beta \mathcal{H}^{\rm{b}} - \beta \tilde{\mathcal{H}}^{\rm{nb}} \right)
\end{split}
\end{equation}
where the Hamiltonian due to the bonding interaction is given by,
\be
\mathcal{H}^b=\sum_{k=1}^{n_{P}}\sum_{s=1}^{N_{P}-1} \frac{3k_BT}{2a^2} \left| {\bf R}_{i,s} - {\bf R}_{i,s+1} \right|^2
\ee
And the Hamiltonian due to the non-bonded interaction is given by,
\be
\mathcal{H}^{\rm{nb}} = \frac{1}{2} \sum\limits_{\alpha = {\rm{P, S, C}}} \sum\limits_{\alpha' \neq \alpha} \int{\rm d}{\bf r}\int{\rm d}{\bf r}' \hat{\phi}_{\alpha} ({\bf r})  u_{\alpha \alpha '}({\bf r}, {\bf r}') \hat{\phi}_{\alpha '} ({\bf r}')
\ee
with $u_{\alpha \alpha '}({\bf r}, {\bf r}')=\chi_{\alpha\alpha'}\delta({\bf r}-{\bf r'})$ and the microscopic number densities of P and S(C) segments at spatial position ${\bf r}$ defined as
\be
\hat{\phi}_{\rm P} ({\bf r}) \equiv \sum_{k=1}^{n_{\rm P}} \sum_{s=1}^{N_P} \delta ({\bf r} - {\bf R}_{P,(k,s)}),
\ee
\be
\hat{\phi}_{\rm S (C)} ({\bf r}) \equiv \sum_{s=1}^{n_{\rm S (C)}} \delta ({\bf r} - {\bf r}_{{\rm S (C)},s}),
\ee
By inserting the identity
\[
1 = \prod_{\alpha={\rm A,B,S,C}} \int \mathscr{D}\phi_{\alpha} \mathscr{D}\omega_{\alpha} \exp \left\{ \int \mathrm{d} \mathbf{r} \omega_{\alpha}(\mathbf{r}) \left[ \phi_{\alpha}(\mathbf{r}) - \hat{\phi}_{\alpha}(\mathbf{r}) \right] \right\},
\]
where $\omega_{\alpha}(\mathbf{r})$ is the purely imaginary conjugate field interacting with species $\alpha $, and applying the saddle point approximation, the standard SCFT equation is given as following,
\begin{equation}
\omega_{\alpha}({\bf r})=\sum_{\alpha'}^{\alpha\neq\alpha'}\chi_{\alpha\alpha'}\phi_{\alpha'}({\bf r})+\xi({\bf r})
\end{equation}
\begin{equation}
\phi_{S}({\bf r})=z_{S}\exp(-\omega_{S}({\bf r}))
\end{equation}
\begin{equation}
\phi_{C}({\bf r})=z_{C}\exp(-\omega_{C}({\bf r}))
\end{equation}
\begin{equation}
\phi_{A}({\bf r})=z_{P}\exp(\omega_{A}({\bf r}))\sum_{s=1}^{N_{A}}q_{s}({\bf r})q_{s}^{*}({\bf r})
\end{equation}
\begin{equation}
\phi_{B}({\bf r})=z_{P}\exp(\omega_{B}({\bf r}))\sum_{s=1}^{N_{B}}q_{s}({\bf r})q_{s}^{*}({\bf r})
\end{equation}
\begin{equation}
\begin{split}
\xi({\bf r}) = & \omega_{C}({\bf r})-\chi_{BC}(1-\phi_{A}({\bf r})-\phi_{S}({\bf r})-\phi_{C}({\bf r}))-\chi_{AC}(1-\phi_{B}({\bf r})-\phi_{S}({\bf r})-\phi_{C}({\bf r})) \\ 
& -\chi_{SC}(1-\phi_{A}({\bf r})-\phi_{B}({\bf r})-\phi_{C}({\bf r}))
\end{split}
\end{equation}
, where $q({\bf r}, s)=\exp(-\omega_{\alpha}({\bf r})) \int {\rm d} {\bf r'} \Phi(\left | {\bf r}-{\bf r'}\right |)q({\bf r'}, s-1)$, $s \leq N_{A}, \alpha=A; s > N_{A}, \alpha=B$ and $q^{*}({\bf r}, N_{P}-s+1)=\exp(-\omega_{\alpha}({\bf r})) \int {\rm d} {\bf r'} \Phi(\left | {\bf r}-{\bf r'}\right |)q({\bf r'}, N_{P}-s+2)$, $s \leq N_{A}, \alpha=B; s > N_{A}, \alpha=A$ are the chain propagators starting from the first and the last segments, respectively. And $\Phi$ is the bond transition factor, $\Phi(\left | {\bf r}-{\bf r'} \right |)=(\frac{3}{2 \pi a^2})^{\frac{3}{2}} \exp(-\frac{3r^2}{2a^2})$. $Q_P$ is the single chain partition function, $Q_P=1/V \int dr \exp(\omega_{A}(r))q(r,1)q^{*}(r,N_{P})$. 

$z_{\alpha}$ is the activity of $\alpha$ component, which is coupled to chemical potential, $\phi_{\alpha}(r)$ is the $\alpha$ component volume fraction at $r$ position, $\chi_{\alpha\alpha'}$ describes the interaction strength between different species, if $\alpha=\alpha'$, $\chi_{\alpha\alpha'}=0$, $\xi$ is the the external potential to ensure the incompressibility condition. Different from common treatment that $\xi$ expression derived by algebra manipulation, we substitute $\phi_{A}(r)+\phi_{B}(r)+\phi_{S}(r)+\phi_{C}(r)=1$ condition into $\omega_{C}(r)$ equation to obtain $\xi$. The reason is that some of $\chi_{\alpha \alpha'}$ being $0$ leads to the incapability to find $\xi$ explicit solution. \par 
Next the system is reduced to one dimension in spherical coordinates by assuming $\psi$ and $\theta$ are constants. The integration of propagator in one dimension can be written as,
\begin{eqnarray}
q(r,s) 
& = & \exp(-\omega_{P}(r)) \int_{0}^{L_{r}}{\rm d}{r'} \int_{0}^{\pi}{\rm d}{\theta'} \int_{0}^{2 \pi}{\rm d}{\psi'} \sin(\theta') {{r'}^2} (\frac{3}{2 \pi a^2})^{\frac{3}{2}} \nonumber \\ & & \exp(-\frac{3}{2 a^2} (r^2 + {r'}^2 - 2 rr' \cos(\theta')) q(r', s) \nonumber \\
& = & (\frac{3}{2\pi a^2})^{\frac{1}{2}} \exp(-\omega_{P}(r)) \int_{0}^{L_{r}} {\rm d}{r'} \frac{r'}{r} (\exp(-\frac{3(r-r')^2}{2a^2}) - \exp(-\frac{3(r+r')^2}{2a^2})) \nonumber \\ & & q(r', s-1)
\end{eqnarray}
Finally, the free energy of the system is,
\begin{eqnarray}
& & \mathcal{H}^{G} [\phi_{A},\phi_{B},\phi_{S},\phi_{C},\omega_{A},\omega_{B},\omega_{S},\omega_{C}]= \nonumber \\
& & \frac{1}{2} \sum_{\alpha =A,B,S,C} \sum_{\alpha' =A,B,S,C} \int d{\bf r} \int d{\bf r'}\phi_{\alpha}({\bf r})u_{\alpha \alpha'}({\bf r}-{\bf r'})\phi_{\alpha'}({\bf r'})- \sum_{\alpha= A,B,S,C}\int d{\bf r} \phi_{\alpha}({\bf r})\omega_{\alpha}({\bf r})  \nonumber \\
& & 
-  z_{P} V Q_{P}[\omega_{A},\omega_{B}] -  z_{S}V Q_{S}[\omega_{S}] - z_{C}V Q_{C}[\omega_{C}]
\end{eqnarray}
\par
The critical point is defined as the $\phi_{P}^{cr}$ where grand potential of the inhomogeneous system equals to the grand potential of homogeneous system (constant solution). The interface of the micelle is decided by the $r^{in}$ where $\phi_{A}(r^{in})=\phi_{B}(r^{in})$. And the aggregation number of the B-block is defined as 
\begin{equation}
N^{agg}_{B} = \int_{0}^{r^{in}}4\pi r^{2}\phi_{B}(r)dr .
\end{equation}

\section{Results and Discussion}
The result section is organized as follows. In the first part, Random phase approximation is used to predict phase instability boundary. In the second part, SCFT calculation provides information about morphology change upon the addition of cosolvents with different quality. In the last part, thermodynamics information about cononsolvency modified micelle is presented, and it is compared with conventional micelle in single solvent.\par
\subsection{Cononsolvency Effects on the Phase Instability} \label{5result_RPA_1}
Random phase approximation is used to figure out the cosolvent excess affinity and solvent composition effect on phase instability boundary, which is spinodal. Excess affinity is defined as the affinity difference of the A-block to solvents and cosolvents, $\Delta{\chi}\equiv \chi_{AS}-\chi_{AC}$.\par
First, we examine the phase diagram of classical A-B/S ternary mixture system. $N_{A}=32$ and $N_{B}=16$ are chosen as A-block (shell) and B-block (core) length, which are the common block length ratios for spherical structure \cite{yin2007simulated}. $\chi_{AB}$ and $\chi_{AS}$ are set as $0.2$ and $0$, respectively. So, A-block and B-block have weak incompatibility, and S is a good solvent for A-block. Figure~\ref{fig:5_figure_1.png} indicates the transition from homogeneous system to inhomogeneous system with the increase of the selectivity of solvents to B-block, which is $\chi_{BS}$. And the system undergoes homogeneous-inhomogeneous-homogeneous transition with the increase of polymer concentration at a given $\chi_{BS}$. In A-B/S ternary mixture system, it is already known that spherical micelle can be formed with high solvent selectivity at diluted polymer concentration \cite{suo2009theoretical}. So, $\chi_{BS}=1.5$ is chosen as the selectivity strength between B-block and solvents, indicated by the red dotted line in figure~\ref{fig:5_figure_1.png}. Therefore, the parameter for "base" system is set.\par

\Figure{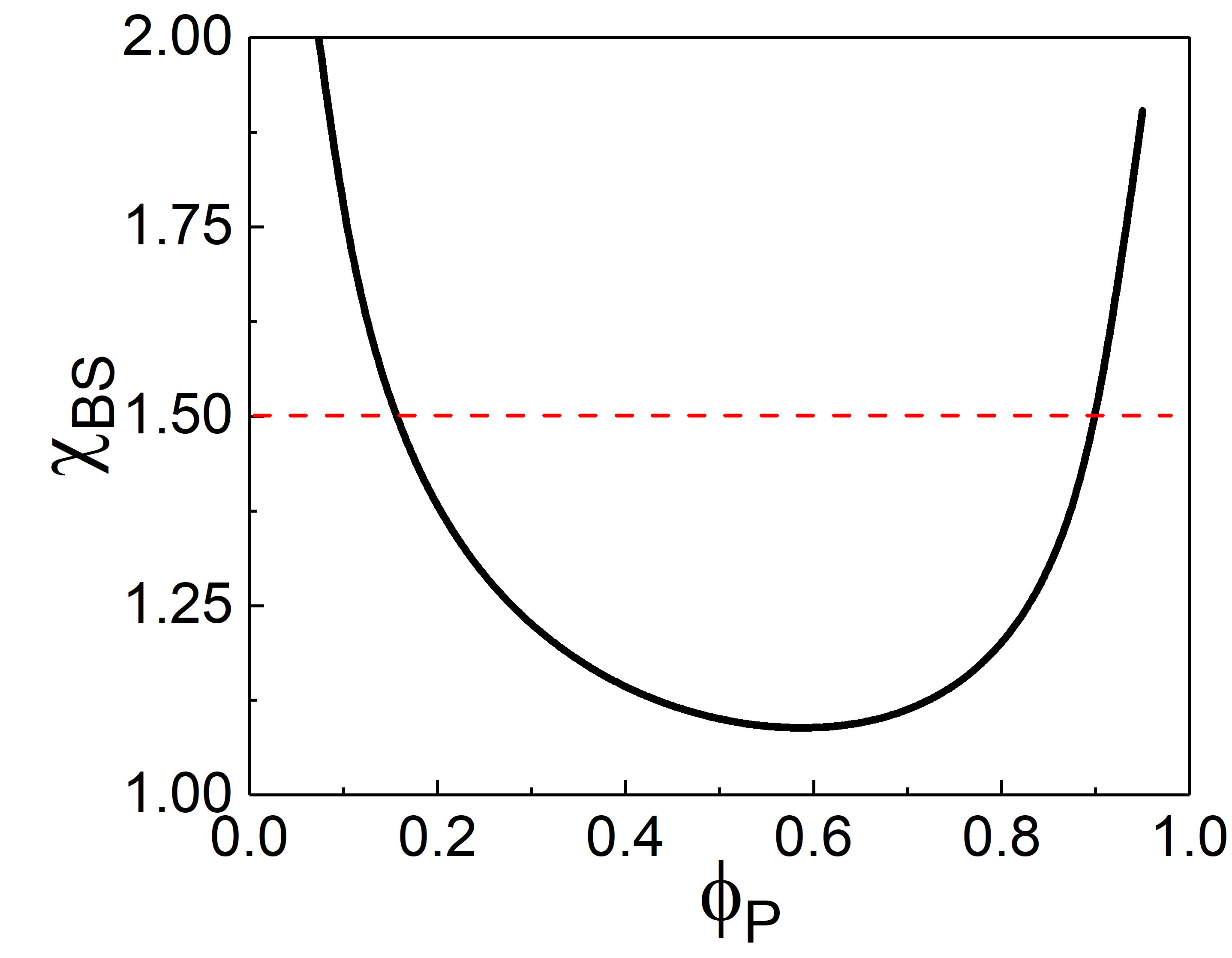}{0.7\linewidth}{Inhomogeneous phase to homogeneous phase transition boundary of A-B diblock copolymer immersed in single solvent.}

After that, cosolvent with different affinity force to A-block is added to the "base" system, corresponding to different $\chi_{AC}$ value, but always keeping $\chi_{BC}=0$. With the addition of different quality cosolvents, figure~\ref{fig:5_figure_2.png} (a) indicates the phase instability boundary change upon varying $\chi_{AC}$ and $x_{C}$. $\phi_{P}^{*}$ is the critical polymer concentration, above which homogeneous state can no longer exist. $x_{C}$ is the cosolvent fraction, which is defined as $x_{C}\equiv\phi_{C}/(\phi_{C}+\phi_{S})$. The continuous increasing of $\phi_{P}^{*}$ with the addition of cosolvents at $\chi_{AC}=0$ system implies that the effective solvent selectivity to B-block(core) is decreasing. In other words, the overall solvent quality is turning better, which can be effectively considered as the decrease of $\chi_{BS}$ in figure~\ref{fig:5_figure_1.png}. And this is similar to PS-PEO (polystyrene–poly(ethylene oxide)) immersed in water/THF(tetrahydrofuran) system. As THF is a good solvent for both blocks, the overall solubility for the polymer is increased \cite{xu1992micellization,seo2002effect}.\par

\Figure{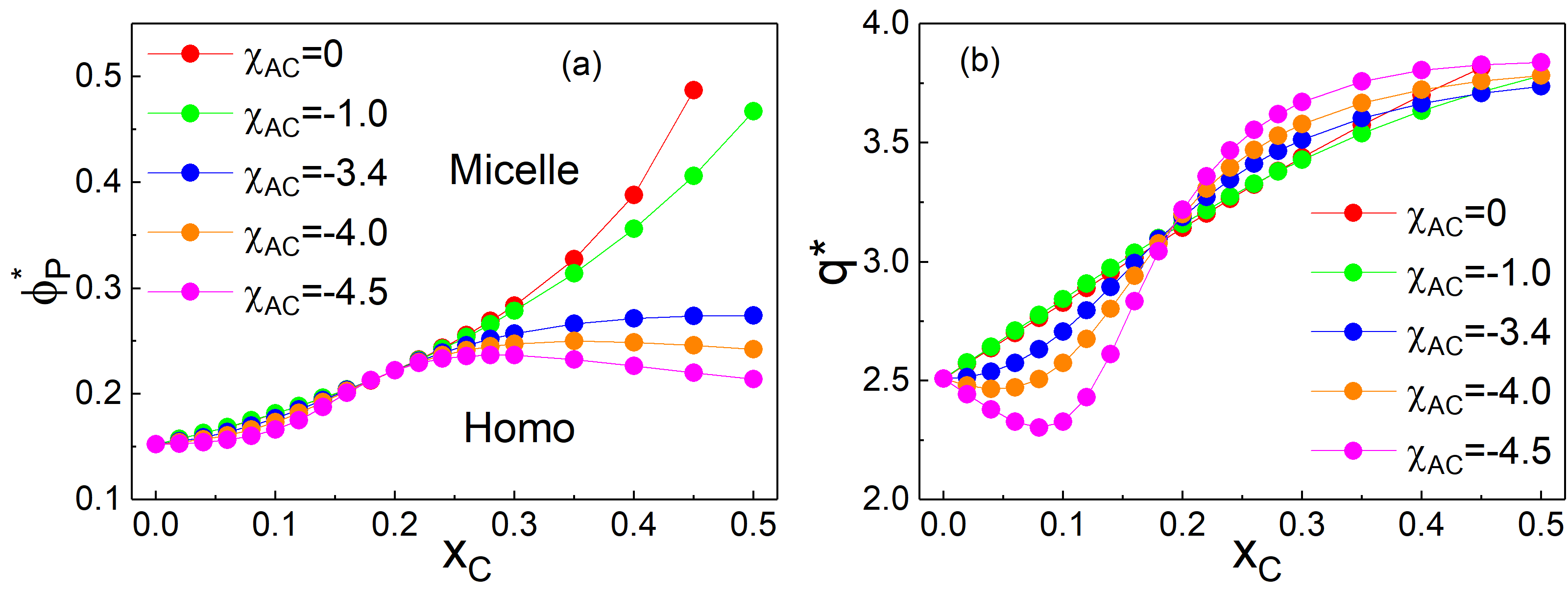}{0.98\linewidth}{(a) Critical polymer concentrations at a certain cosolvent fraction in different cosolvent excess affinity systems calculated by RPA model. (b) Diverging wave vector length plotted against cosolvent fraction with different cosolvent excess affinity in RPA calculation.}

As $\chi_{AC}$ becomes more negative, the cosolvent quality turns better for the A-block. Because cosolvent is also a good solvent for B-block(core), the increase of the critical polymer concentration should be expected at the same cosolvent fraction, when the $\chi_{AC}$ value was decreased. As the overall solubility for polymer should be increased, the inhomogeneous region in phase diagram should become smaller. But we see the contrary result in figure~\ref{fig:5_figure_2.png} (a) that $\phi_{P}^{*}$ is decreased at high cosolvent fraction with smaller $\chi_{AC}$. The addition of the other type of solvents with better quality increases the effective solvent selectivity and expands the inhomogeneous window. This counter-intuitive behavior can be ascribed to cononsolvency effect, which worsens the effective solvent quality, though only the micelle shell exhibits its effect. The similar dependence of critical polymer concentration on excess affinity predicted by Flory-Huggins theory is also reported for homopolymer exhibiting cononsolvency system. The increase of excess affinity will promote the phase separation during the occurrence of cononsolvencye effect. Figure~\ref{fig:5_figure_2.png} (b) indicates the change of diverging wave vector length as a function of cosolvent fraction. $q^{*}$ being close to $0$ or finite value suggests whether the phase separation is macro-scale or micro-scale \cite{leibler1980theory}. We can see that figure~\ref{fig:5_figure_2.png} (b) suggests micro-phase separation occurring in this system.  The development of the minimum with the improvement of the cosolvent quality indicates the morphology variation, but RPA is not capable of providing more details, especially system micro-structure information. To investigate the puzzling critical polymer concentration variation behaviors and to verify the formation of the micelle structure, thermodynamics and component distribution results are needed. Therefore, SCFT calculation is applied. \par
\subsection{Cononsolvency Effects on the Micellar Morphology} \label{5result_SCFT_morphology}
First, same as RPA study, critical micelle concentration and aggregation number are plotted against cosolvent fraction at different excess affinity, corresponding to different $\chi_{AC}$ value, as shown by figure~\ref{fig:5_figure_3.png}. Phase boundary is defined as the point where the grand potential difference between homogeneous system and micelle system equals to zero, so, the curve represents the binodal. It can be observed that phase boundary predicted by SCFT shows the qualitatively same trend with RPA results. The increase of cosolvent fraction at the same $\chi_{AC}$ decreases the CMC. But CMC tends to be decreased with the improvement of cosolvent quality at the same cosolvent fraction, suggesting the happening of cononsolvency effect. Though CMC of A-B/S/C system is not measured directly in experiments, the decrease of the cloud point with the increase of the cosolvent fraction can be observed, indicating the occurrence of cononsolvency effect \cite{dalkas2006control,kyriakos2014cononsolvency}. \par

\Figure{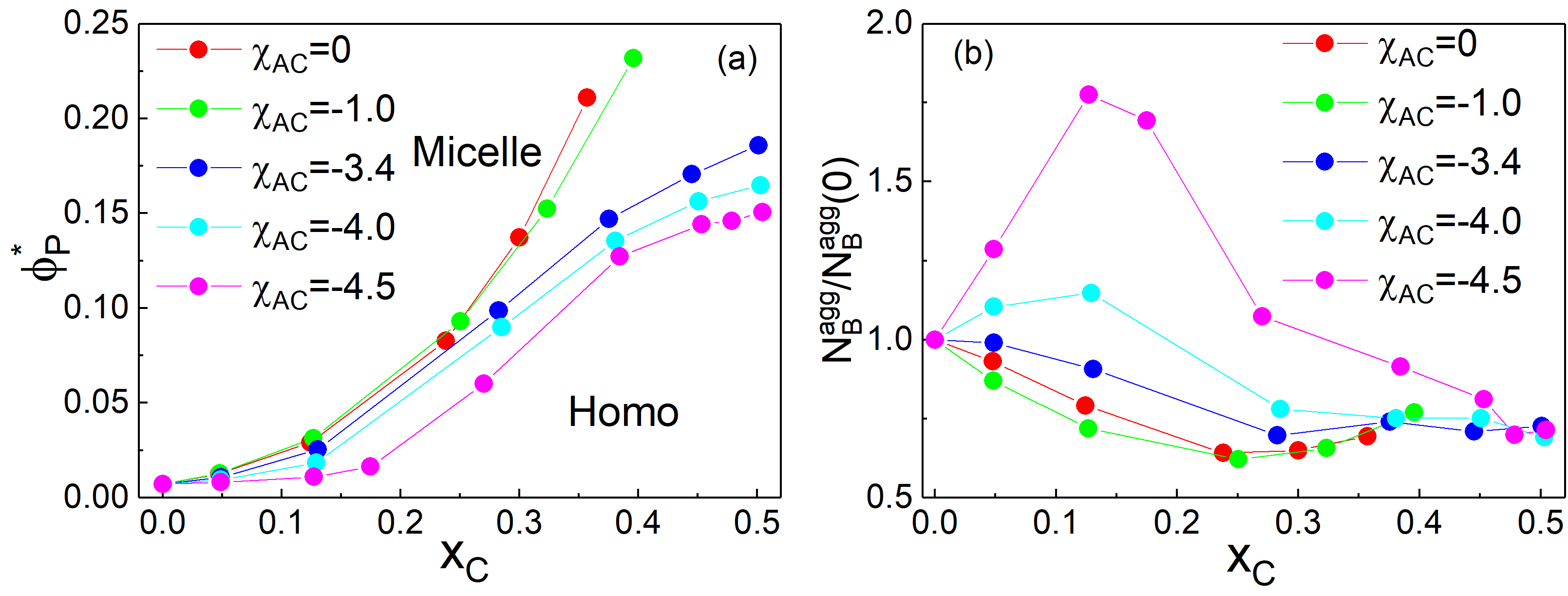}{0.98\linewidth}{(a) The binodal boundary of A-B/S/C system calculated by SCFT model in different cosolvent excess affinity systems. (b) Micelle aggregation numbers normalized by A-B/S system plotted against cosolvent fraction.}

The aggregation number of B-block segments is a direct indicator of micelle morphology. The value of the reciprocal of $q^{*}$ in RPA calculation indicates the micelle size change. The larger $q^{*}$ becomes, the smaller the micelle size is. So, it can be found that the SCFT aggregation number variation is qualitatively consistent with RPA $q^{*}$ change. In both RPA and SCFT results, the maximum point of micelle size begins to develop with the increase of the excess affinity, as cononsolvency phenomenon becomes more evident. In $\chi_{AC}=0$ and $\chi_{AC}=-1$ system, aggregation number almost monotonically decreases with the addition of cosolvents, though a little uptick tail can be observed. Compared with pure solvent system, the micelle size is still shrinking at large cosolvent fraction. This is what usually is observed in conventional micelle system upon the addition of the secondary good solvent when cononsolvency effect does not come into the picture. The micelle core becomes soften due to the decrease of the effective solvent selectivity, and correspondingly, more solvents and cosolvents penetrate into the core. Therefore, number of polymer segments inside the core becomes less \cite{seo2002effect,schaeffel2014molecular}. The "abnormal" behavior caused by cononsolvency begins to emerge if A-block - cosolvent affinity strength is further increased, meaning the decrease of $\chi_{AC}$ value. The aggregation number shows the increasing trend with the decrease of $\chi_{AC}$. A maximum point begins to develop at cosolvent fraction equal to $0.12$. Especially in $\chi_{AC}=-4.5$ system, an evident peak has shown up, suggesting the significant increase of B-block segments number inside the core. Similar behaviors have been reported in experiments. The final aggregate size of polystyrene-b-poly(N-isopropylacrylamide)(PS-b-PNIPAM) immersed in water/methanol mixture increases with the addition of methanol(cosolvent) \cite{kyriakos2014cononsolvency}. But volume fraction mixing ratio of water/methanol only increases to $80:20$ in their study. In general, cononsolvency effect should begin to lessen from the aspect of both chain conformation and phase behavior after a certain mixing ratio if we keep increasing the cosolvent fraction, which can be extrapolated from the homopolymer system \cite{zhang2001reentrant,schild1991cononsolvency,costa2002phase}. The same tendency can be observed in micelle system in our study. The aggregation number begins to drop if we keep adding the cosolvent after the maximum point, indicating the diminishing cononsolvency effect. At last, the micelle in all systems with different $\chi_{AC}$ value goes to a similar size, which can also be reflected in density file plots.\par
The density profile plots at different $\chi_{AC}$ and $x_{C}$ can provide a straightforward description about the morphology change. In figure~\ref{fig:5_figure_4.png}, y-axis is the component fraction at the position $r$. First, $\chi_{AC}=0$ system structure change with the increase of the cosolvent fraction is shown. It can be observed that the degree of aggregation of B-block segments exhibits evident decrease with $x_{C}$ increasing due to the improvement of overall solvent quality. It has been shown that the decrease of the aggregation is caused by the decrease of the corona-core interfacial tension as the cosolvent is good solvent for both blocks \cite{seo2002effect,kelley2011structural}.  Accordingly, the micelle core becomes soften, or in other words, it becomes more accessible for both solvents and cosolvents. It can be observed in the plot that solvents and cosolvents show the significant enrichment inside the core with $x_{C}$ increasing, and they take up polymer segments' space, which can also be deduced by previous aggregation number plot. Next, it can be seen that the density profile change affected by cononsolvency is quite different. Although the penetration of solvents and cosolvents into the micelle core can be observed in both type of systems, the polymer distribution is totally different. The shell, which is A-block, obviously exhibits the conformational nonmonotonic change. The extent of collapse is greatly increased at $x_{C}=0.13$ point due to cononsolvency effect, which is also reported by experiments \cite{ko2021co}. Meanwhile, the aggregation of the shell also causes the penetration of A-block into the micelle core, so, the number of A segments inside the core is increased by comparing the density profile at $x_{C}=0.13$ with $\chi_{AC}=0$ system, which can also be observed in the experiments \cite{wang2008investigation}. The collapse of the micelle shell may play the role to stabilize the micelle, and in further, promote the formation of larger aggregates \cite{kyriakos2016quantifying}. Clearly, the increase of the micelle size can be observed at $x_{C}=0.13$, $\chi_{AC}=-4.5$ system, which is also suggested by previous aggregation number plot. Thermodynamics analysis may provide more information about the structure change, which will be discussed in the next part. When we keep adding the cosolvent to the system, it can be seen that the shell becomes extended, similar to cononsolvency induced conformational variation in homopolymer system. Therefore, the system behavior gradually recovers with $\chi_{AC}=0$ system as cononsolvency effect is vanishing, which may confirm the argument that the collapsed shell can stabilize the micelle.\par

\Figure{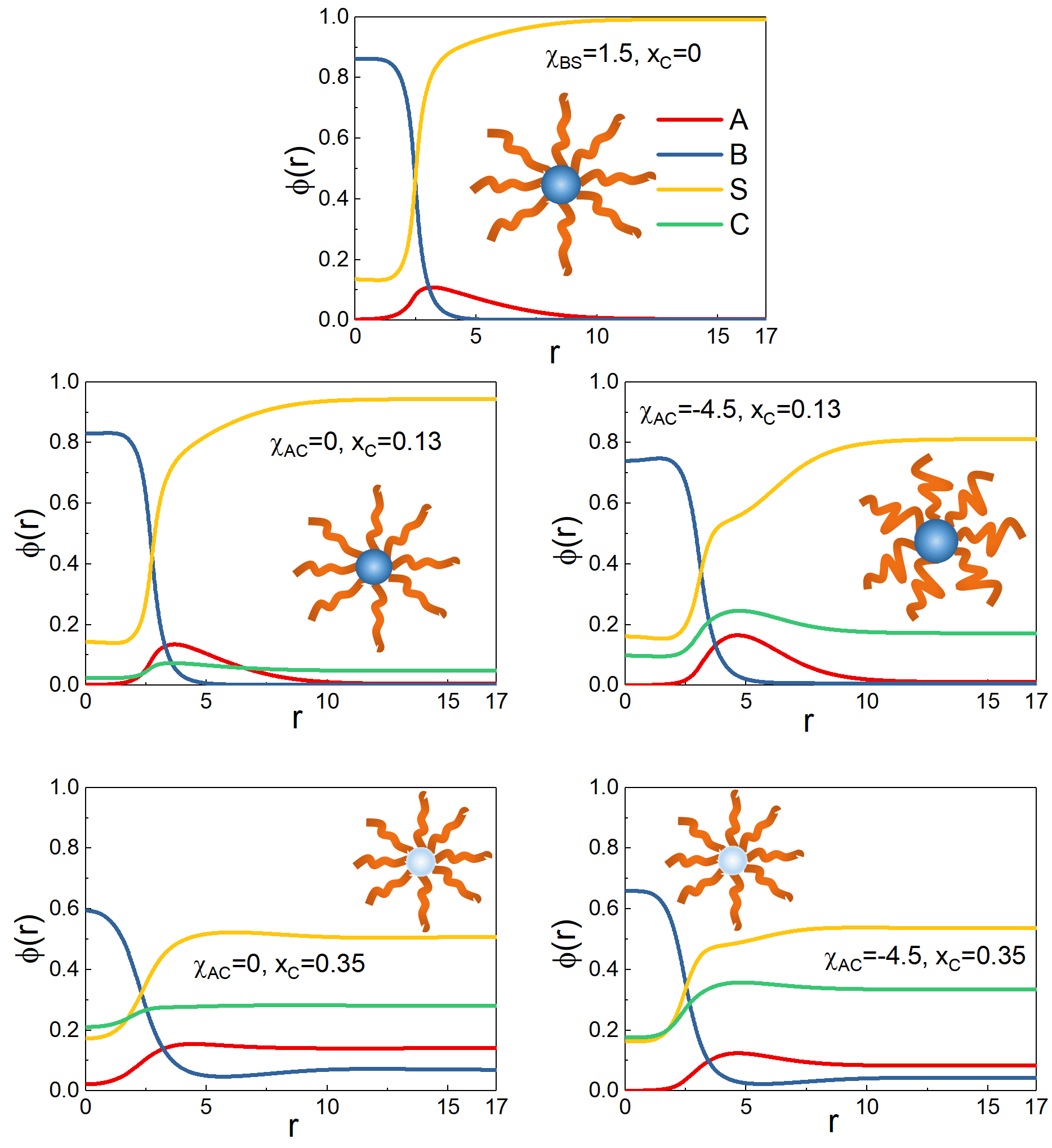}{0.98\linewidth}{The evolution of density profiles at $\chi_{AC}=0$ and $\chi_{AC}=-4.5$ with the increase of the cosolvent fraction. And the corresponding illustration plot of micelle morphology is shown in the inset.}
\subsection{Cononsolvency Effects on the Micelle thermodynamics - Different Driving Force} \label{5result_}

The occurrence of the cononsolvency effect can be differentiated more clearly by the chemical potential plot. Figure~\ref{fig:5_figure_5.png} shows the chemical potential of the polymer as a function of cosolvent fraction at the binodal boundary, where the grand potential of homogeneous system equals to the grand potential of micellar system. It can be found that the system without cononsolvency effect shows the increasing polymer chemical potential with the addition of cosolvents, because polymer concentration (binodal boundary) is also increasing with the addition of the cosolvent in $\chi_{AC}=0$ and $\chi_{AC}=-1$ system as it is shown in figure~\ref{fig:5_figure_3.png}. But the system with cononsolvency occurring shows the decreasing trend of polymer chemical potential with cosolvent fraction increasing. The decrease of the chemical potential usually suggests the decrease of the polymer concentration but that is not the case as it is shown in figure~\ref{fig:5_figure_3.png}. The reason is more likely the fortification of the micelle structure due to cononsolvency effect which can be explained better in figure~\ref{fig:5_figure_6.png}, which is thermodynamics driving force analysis.\par

\Figure{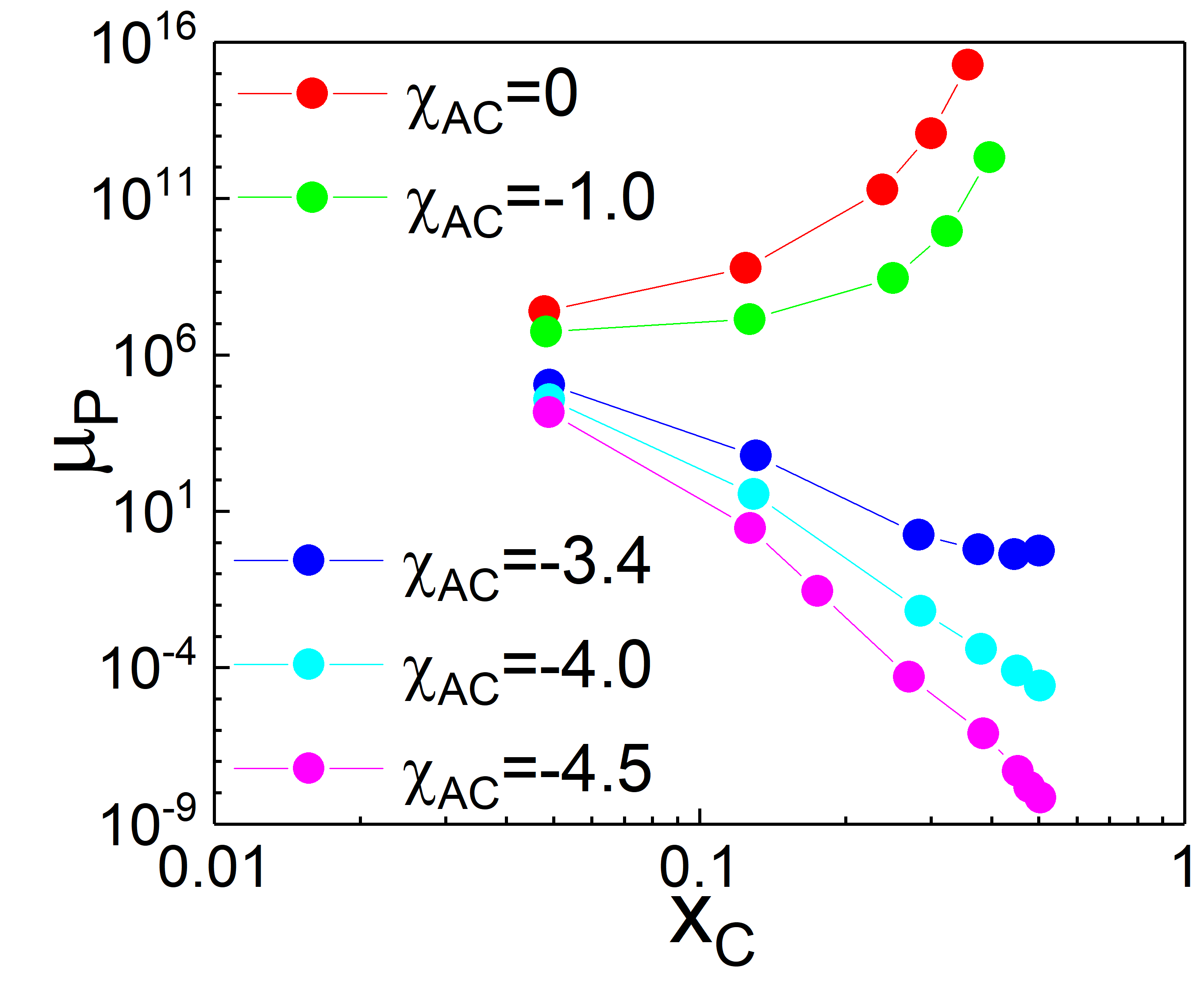}{0.7\linewidth}{Polymer chemical potential with different cosolvent excess affinity as a function of cosolvent fraction.}

Figure~\ref{fig:5_figure_6.png} (a) and (b) show the free energy component difference in A-B/S system with $\chi_{BS}=1.5$ and $\chi_{BS}=1.3$. In large selectivity system, the micelle formation is driven by entropy when $\phi_{P}$ is at extremely dilute region. The system entropy loss compensates for the enthalpy gain because polymer chains only take up a quite small volume of the system. The increase of polymer concentration will result in the different driving forces. The transfer of the hydrophobic block into the micelle core results in the reduction of B-block - solvent hydrophobic energy, and finally, the stable core-solvent interface is formed \cite{nagarajan1989block}. At small selectivity system, critical polymer concentration cannot reach the dilute region, so it cannot observe the entropy driven stage. The micelle formation is driven only by the decrease of B-S interaction caused by the polymer aggregation. Figure~\ref{fig:5_figure_6.png} (c) and (d) compare the free energy component difference in A-B/S/C system with $x_{C}=0.13$ at $\chi_{AC}=0$ and $\chi_{AC}=-4.5$ system. In $\chi_{AC}=0$ system, the micelle formation is driven by the decrease of B-S repulsive energy, which is similar to A-B/S ternary mixture system. It tells that the driving force is still the effective solvent selectivity. The addition of the cosolvent does not change the mechanism of the micellization. But in $\chi_{AC}=-4.5$ system, different trends can be observed. Strong A-C favorable interaction can compensate for the entropy gain when the polymer concentration increases. The highly aggregated micelle shell becomes the protection shell to isolate the micelle from the outside components, as the A-block segments enclose the micelle core tightly, which is shown by previous density profile plot. The shell can stabilize the micelle structure and promotes the formation of large aggregates due to A-C attractive interactions, and moreover, causing the expansion of the micelle size \cite{kyriakos2016quantifying}. A slight increasing of $\Delta E_{BS}$ can be observed at the low polymer concentration, but after a certain point, $\Delta E_{BS}$ curve will go down. This behavior is similar to trends in $\chi_{BS}=1.5$ system, corresponding to figure~\ref{fig:5_figure_6.png} (a), caused by the extremely low polymer concentration. From the free energy component analysis, it can be concluded that A-C favorable interaction can significantly change the micelle behavior. The addition of the secondary type solvents with better quality will result in the deterioration of overall solvent mixture quality. Hence, the micelle formation is promoted. But when cosolvent fraction reaches a certain value, the solvent mixture quality turns good again like vanishing cononsolvency effect in all of types of systems. And micelle structure will be finally undermined due to the penetration of solvents and cosolvents into the micelle core. Thus, the whole process of micelle expansion is dominated by cononsolvency effect. Overall, the addition of good cosolvents drives the collapse of A-block corona and in further modifies micelle properties.\par

\Figure{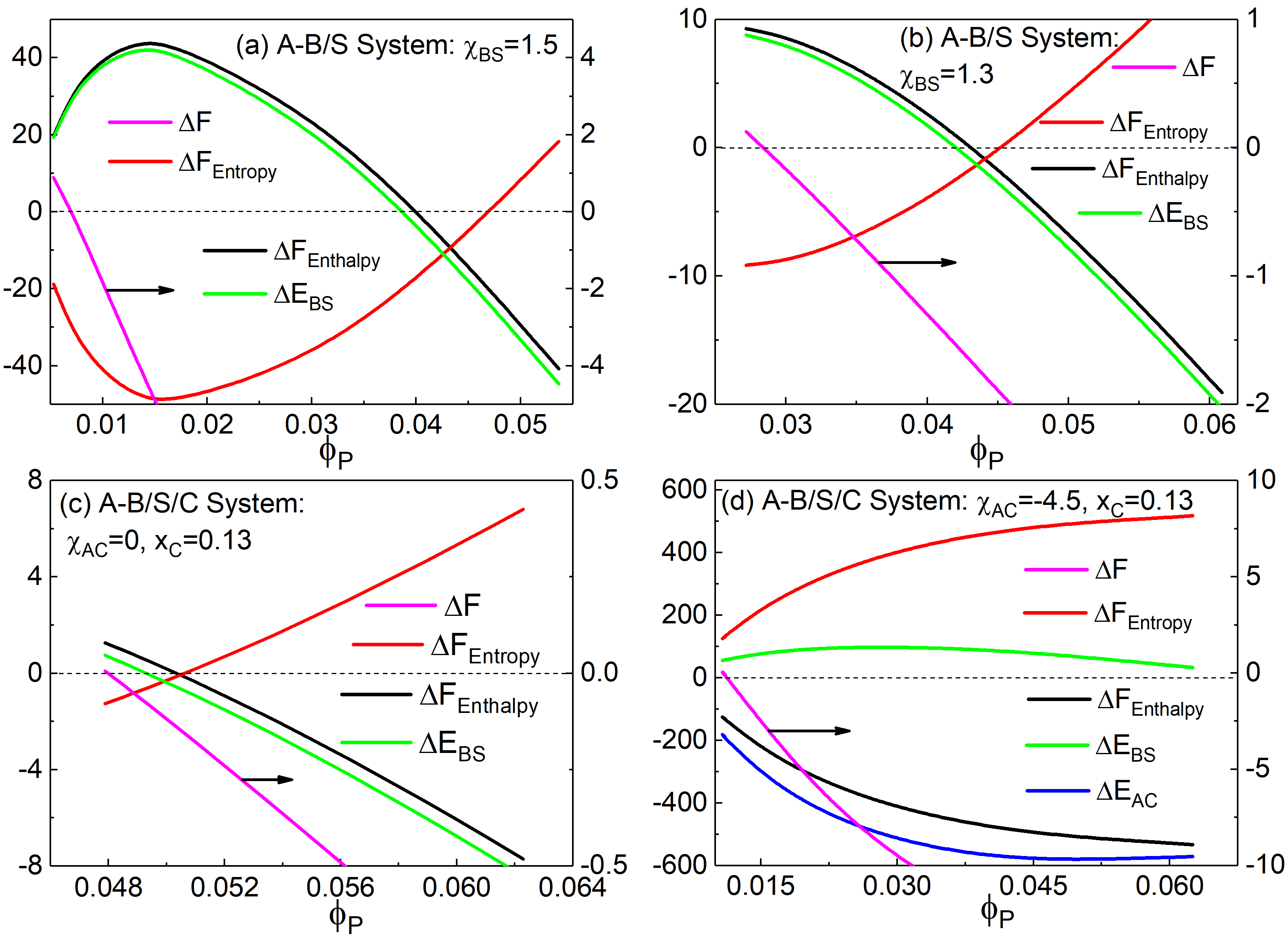}{0.98\linewidth}{Difference of free energy components ($F_{X}$(micelle) - $F_{X}$(homo) plotted against the polymer concentration in A-B/S system with (a) $\chi_{BS}=1.5$  and (b) $\chi_{BS}=1.3$, and A-B/S/C system with (c) $\chi_{AC}=0, x_{C}=0.13$ and (d) $\chi_{AC}=-4.5, x_{C}=0.13$. }

\section{Conclusion}
The critical micelle concentration and micelle size predicted by RPA are qualitatively agreed with SCFT results. The occurrence of cononsolvency effect can expand the micelle size and decrease the critical micelle concentration, compared with non-cononsolvency system. The above behaviors can be ascribed to micelle structure change. The micelle shell exhibits extension-collapse-extension transition due to cononsolvency effect. The shrinkage of the micelle shell can stabilize the micelle and promote the aggregation, and finally result in the micelle size growth. The occurrence of cononsolvency effect can be better indicated by chemical potential plots. It shows that cononsolvency effect will come into the picture only when $\chi_{AC}$ is smaller than $-3.4$. By analyzing the free energy component plots, it can be found that the decrease of shell-block - cosolvent interaction plays the major role to minimize the total free energy, which is different from conventional micelle in single solvent. The shell-block - cosolvent attractive interaction and core-block - solvent repulsive interaction both drive the micelle formation when cononsolvency comes into effect.

\newpage
\bibliographystyle{unsrt}
\bibliography{citation}

\end{document}